\documentclass[a4paper]{panl}
\usepackage{cite}
\usepackage{wrapfig}
\usepackage{graphicx}
\usepackage{amssymb}
\usepackage{amsfonts}
\usepackage{amsmath}
\usepackage{longtable}
\usepackage{rotating}
\usepackage{lscape}
\usepackage{epsfig}
\usepackage{multirow}

\usepackage{url}
\usepackage{hyperref}

\originalTeX
\begin{document}
\issuearea{Physics of Elementary Particles and Atomic Nuclei. Experiment}

\title{Evaluation of the inclusive electron scattering observables in the resonance region from the experimental data \\ Оценка наблюдаемых инклюзивного рассеяния электронов в резонансной области на основе экспериментальных данных}
\maketitle
\authors{A.\,A.\,Golubenko$^{a,}$\footnote{E-mail: aa.golubenko@physics.msu.ru}, 
V.\,V.\,Chesnokov$^{a,}$, B.\,S.\,Ishkhanov$^{a,}$, V.\,I.\,Mokeev$^{b,}$}
\from{$^{a}$\,Skobeltsyn Nuclear Physics Institute at Moscow State University, 119899 Moscow, Russia}
\vspace{-3mm}
\from{$^{b}$\,Thomas Jefferson National Accelerator Facility, Newport News, Virginia 23606, USA}

\begin{abstract}
Представлен метод оценки наблюдаемых инклюзивного рассеяния электронов в резонансной области на основе экспериментальных данных. Управляемая через web-интерфейс программа позволяет пользователю рассчитать сечения и структурные функции $F_{1}$ и $F_{2}$ для инклюзивного рассеяния электронов на протоне при инвариантных массах конечного состояния 1.0 ГэВ $<$ W $<$ 4.0 ГэВ и виртуальностях фотона 0.5 ГэВ$^2$ $<$ Q$^2$ $<$ 7.0 ГэВ$^2$. Предложенный подход представляет интерес для анализа экспериментальных данных при исследовании структуры адронов в электромагнитных процессах.\
\vspace{0.2cm}

The method for evaluation of inclusive electron scattering observables in the resonance region from the experimental data is presented. The computer codes in the web page allow user to evaluate inclusive electron scattering, virtual photon-proton cross sections and the $F_{1}$ and $F_{2}$ 
structure functions at the invariant masses of the final hadrons 1.0 GeV $<$ W $<$ 4.0 GeV and at photon virtualities 0.5 GeV$^2$ $<$ Q$^2$ $<$ 7.0 GeV$^2$. The developed approach is of interest for analyses of the experimental data on exploration of the hadron structure in the experiments with electromagnetic probes.  
\end{abstract}
\vspace*{6pt}


\section{Introduction}

The studies of exclusive meson electroproduction off protons with the CLAS detector at Jefferson Lab extended considerably the knowledge on the structure of the ground and excited nucleon states \cite{BurkertRoberts,aznaryanBurkert}. Currently they are the only source of the information on many facets of the strong QCD dynamics in generation of the excited nucleon states of different quantum numbers with distinctively different structural features. The CLAS detector has produced the dominant part of the available world data on all meson electroproduction channels off the nucleon relevant in the resonance region for $Q^2$ up to 5.0~GeV$^2$. The numerical values of all observables measured with the CLAS detector are stored in the CLAS Physics Database~\cite{clasdb}. Analyses of these experimental data within the framework of the reaction models developed by the CLAS collaboration \cite{Az09,park15,Den07,Mo09,Mo12,Mo16} provided the first and only available in the world information on electroexcitation amplitudes $A_{1/2}(Q^2)$, $A_{3/2}(Q^2)$, and $S_{1/2}(Q^2)$ ($\gamma_{v}pN^*$-electrocouplings) of most excited nucleon states in the mass range up to 1.8 GeV and at photon virtualities $Q^2$ up to 5.0 GeV$^2$ \cite{Mokeev17}.The numerical values of the $\gamma_{v}pN^*$- electrocouplings from CLAS and the computer codes for their interpolation/extrapolation at $Q^2$ $<$ 5.0 GeV$^2$ can be found in the web pages ~\cite{mokeev-web,isupov-web}. The CLAS results on the photocouplings of most excited nucleon states in the mass range up to 2.0 GeV have recently become available from the exclusive $\pi^+\pi^-p$ photoproduction off protons data \cite{Gol19}. In the near term future, electrocouplings of most excited nucleon states in the mass range up to 2.0 GeV will become available at $Q^2$ up to 5.0 GeV$^2$ from analyses of the CLAS exclusive meson electroproduction data \cite{Is17,Triv18,Fed18}.

Physics analyses of these results revealed the structure of the excited nucleon states as a complex interplay between inner core of three dressed quarks and external meson-baryon cloud ~\cite{BurkertRoberts,aznaryanBurkert,Mo12,Mo16}. Analyses of the CLAS results on $\gamma_{v}pN^*$-electrocouplings within continuum QCD Dyson-Schwinger Equation (DSE) approach conclusively demonstrated the capability of getting insight into strong QCD dynamics underlying the generation of the dominant part of hadron mass from combined studies of the experimental results on the nucleon elastic form factors and electrocouplings of different excited nucleons with distinctively different structural features \cite{Seg14,Seg15,Mo16}. Insight to the mechanisms underlying the hadron mass generation was confirmed in the studies of the CLAS results on the nucleon elastic form factors and $\gamma_{v}pN^*$-electrocouplings within independent framework of the novel relativistic quark model which incorporates the momentum dependence of constituent quark mass and the modeling of meson-baryon cloud \cite{aznurayan12,aznurayan15,aznurayan17}.

Detailed information on the electrocouplings of most excited nucleon states from the CLAS offers the new opportunities in exploration of the ground nucleon structure from the inclusive electron scattering data. The data on inclusive electron scattering off protons described in terms of well-known structure functions $F_{1}(x_{B},Q^2)$ and $F_{2}(x_{B},Q^2)$ offer an access to the parton distribution functions (PDF) in the ground nucleons $f_{i}(x_{B},Q^2)$ (index $i$ stands for all flavor current quarks/anti-quarks and gauge gluons) \cite{Ac16}. The PDFs determine the probability to find $i$th-parton with fraction x of the total nucleon momentum. They are defined on the light front and evolve with the distance scale determined by the photon virtuality $Q^2$. The behavior of the parton distributions at large values of $x_{B}$ corresponded to the nucleon resonance excitation region is the subject of a special interest. At large $x_{B}$ the perturbative-QCD (pQCD) prediction on behavior of parton distributions become possible \cite{Ac16}. Furthermore, the development of novel pseudo- and quasi-PDF concepts \cite{Lin18,Qiu18,Rad19} makes it possible to relate the non-perturbative behavior of the PDFs to the QCD Lagrangian by employing lQCD evaluations for the pseudo- and quasi-PDF and taking the appropriate limit for the transformation of these approximations for PDF into the accessible from the experimental data PDFs.

So far, the PDFs were evaluated either outside the resonance excitation region, or, under theoretical assumption on quark-hadron duality \cite{Mel05,Mal09}, within the resonance excitation region. The CLAS results on $\gamma_{v}pN^*$-electrocouplings for the first time make it possible to evaluate the resonant contributions into the $F_{1}(x_{B},Q^2)$ and $F_{2}(x_{B},Q^2)$ inclusive electron scattering structure functions based on the results of experiment on the nucleon resonance electroexcitation amplitudes and improve in this way the information on the parton distributions in the ground nucleon states at large $x_{B}$ within the resonance excitation region.

In this proceeding we present the approach for evaluation of the inclusive electron scattering observables, inclusive electron scattering, virtual-photon-proton cross sections, and the $F_{1}(x_{B},Q^2)$ and $F_{2}(x_{B},Q^2)$ inclusive structure functions at 1.07 GeV $<$ W $<$ 4.0 GeV and 0.5 GeV$^2$ $<$ $Q^2$ $<$ 7.0 GeV$^2$ from the available experimental results. The aforementioned observables can be evaluated in the real time in the web-page \cite{Chesn19} with the numerical or graphical outcome upon the user request. Reliable evaluation of the inclusive electron scattering observables in the resonance excitation region represents the first step towards accessing the ground nucleon PDF at $x_{B}$ in the resonance region and accounting for the resonance contributions based on the experimental results on $\gamma_{v}pN^*$-electrocouplings.

First experiments with the CLAS12 detector already started successfully in February 2018. The experimental program with the CLAS12 detector is focused on 3D imaging of the ground nucleon states from the data of Deeply Virtual Compton Scattering (DVCS), Deeply Virtual Meson Production (DVMP) and the measurements of the Transverse Momentum Distributions (TMD), extension of the results on $\gamma_{v}pN^*$-electrocouplings towards highest $Q^2$ ever achieved in exclusive reactions up to 12 
GeV$^2$, search for new state of baryon matter, the so-called hybrid baryons with glue as an active structural component, meson spectrum exploration through multi-meson final state production by the quasi-real photons, studies of strange baryons and J/$\psi$ production \cite{Bu19}. The data on inclusive electron scattering observables offer an important benchmark for all the aforementioned experiments allowing us to validate normalization of semi-inclusive and fully exclusive cross sections as well as credible evaluation for the scattered electron detection efficiency.

\section{Method for evaluation of the inclusive electron scattering of observables}

\begin{figure}
\centering
    \includegraphics[width=6 cm]{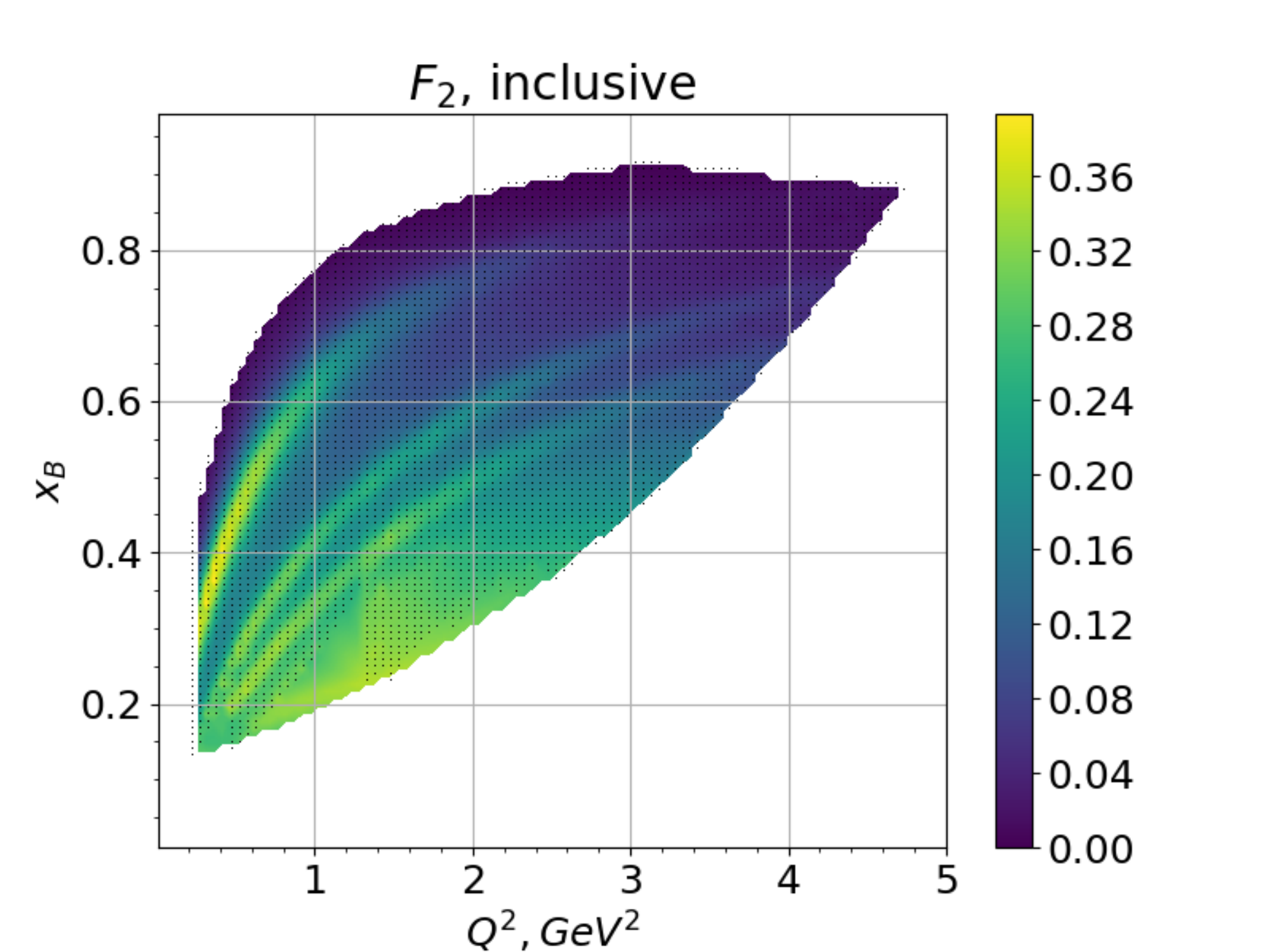}
    \includegraphics[width=6 cm]{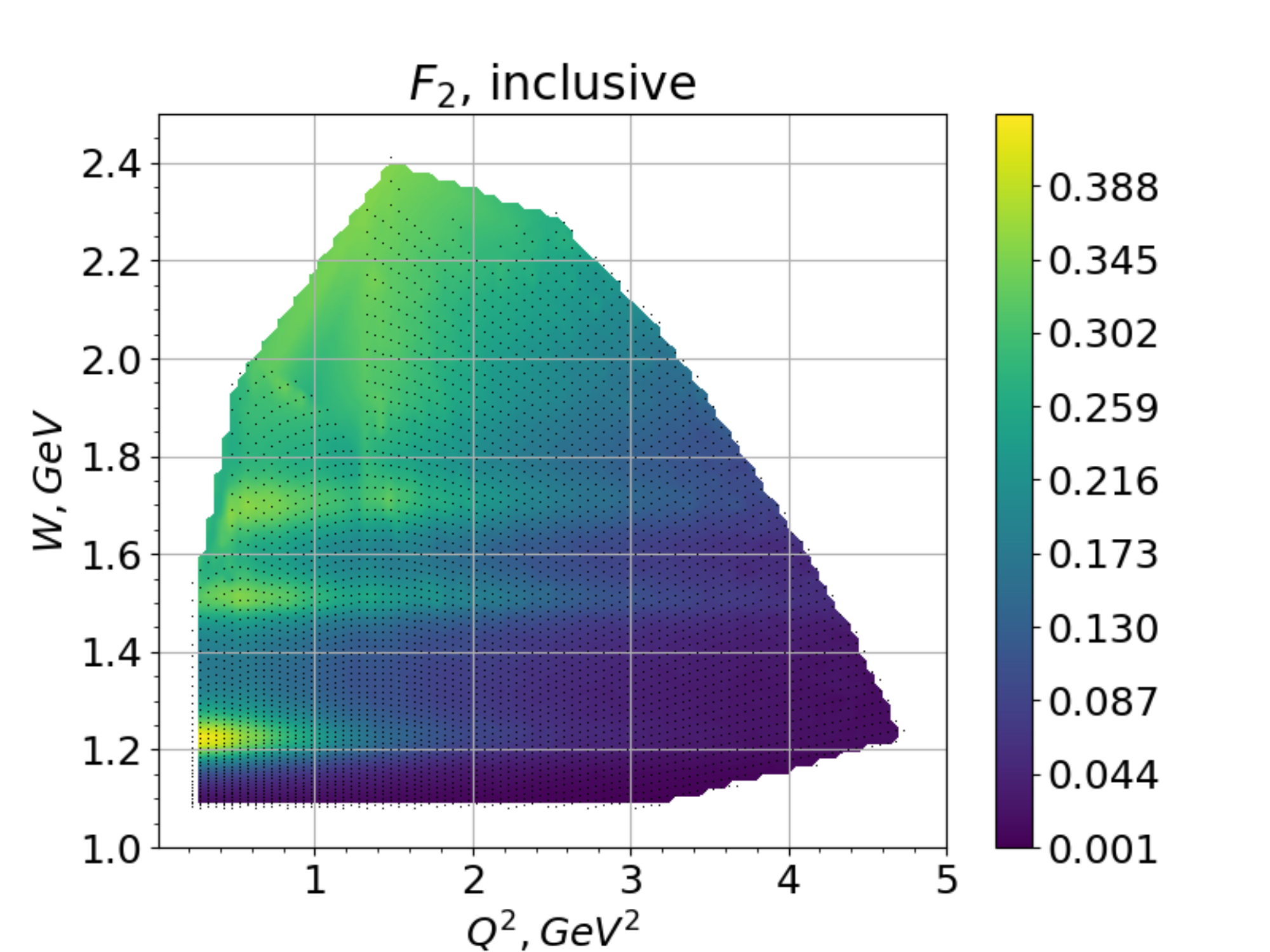}
    \caption{The kinematic coverage for the CLAS data~\cite{osip} on the $F_{2}$ inclusive structure function Q$^2$ vs $x_{B}$ (left) and 
Q$^2$ vs W (right). The data points are shown by black dots. Three strips clearly seen in the 2D-distributions correspond to the first-, second-, and third-resonance regions. The structure function increase at small $x_{B}$ (large $W$) corresponds to the transition into the deep inelastic scattering (DIS) regime.}
    \label{fig1}
\end{figure}

We develop the approach for evaluation of the $F_{1}(x_{B},Q^2)$ and $F_{2}(x_{B},Q^2)$ inclusive structure functions, inclusive electron scattering and virtual photon-proton cross sections from the available experimental results on the structure functions from CLAS and worldwide in the kinematic area of the invariant masses of the final hadrons 1.07 GeV $<$ W $<$ 4.0 GeV and the photon virtualities  0.5 GeV$^2$ $<$ Q$^2$ $<$ 7.0 GeV$^2$. We are using the CLAS data 
\cite{osip} on $F_{2}(x_{B},Q^2)$ inclusive structure function in the resonance excitation region with the kinematic coverage shown in Fig.~\ref{fig1}. These data can be found in the CLAS Physics Data Base \cite{clasdb}.

Outside the CLAS data kinematics coverage and in the entire $W$-range at $Q^2$ $>$ 1.5 GeV$^2$, parameterization \cite{bost} of the measured worldwide inclusive structure functions is employed as the estimate for the experimental results. The CLAS data offer certain advantage in the resonance excitation region. Because of the almost 4$\pi$ coverage, the CLAS detector is capable of obtaining $F_{2}(x_{B},Q^2)$ inclusive stricture function within much larger range of W ($x_{B}$) in comparison with achievable with electron spectrometers of small angular acceptance for the scattered electrons. This advantage is of particular importance for the evaluation of the inclusive electron scattering observables in the resonance excitation region with pronounced resonance structures clearly seen in Fig~\ref{fig1}.

The interpolation of the experimental results is done independently in each bin of W. The Q$^2$-evolution of the inclusive structure functions was parameterized by employing Q$^2$-dependencies expected from the operator product expansion for the moments of inclusive structure functions:
\begin{equation}\label{OPE}
\begin{aligned}
F_{1}(W,Q^2)=C_{0,1}(W)+ \frac{C_{1,1}(W)}{Q^2}+ \frac{C_{2,1}(W)}{Q^4} + ... \\
F_{2}(W,Q^2)=C_{0,2}(W)+ \frac{C_{1,2}(W)}{Q^2}+ \frac{C_{2,2}(W)}{Q^4} + ...
\end{aligned}
\end{equation}

The parameters $C_{i,j}$ (i=0,1,2,j=1,2) are obtained from the fit of the experimental results on Q$^2$-evolution of the $F_{1}(W,Q^2)$ and $F_{2}(W,Q^2)$ inclusive structure functions from the CLAS data \cite{osip} or by combining the CLAS data and the parameterization \cite{bost} of the world data. The structure function $F_{1}(W,Q^2)$ was computed from the structure function $F_{2}(W,Q^2)$ assuming parameterization \cite{Ric99} for the longitudinal over transverse 
$\sigma_{L}/\sigma_{T}$-cross section ratio used in the extraction of the structure function $F_{2}(W,Q^2)$ from the CLAS inclusive electron scattering cross section \cite{osip}.

The computed in this way the $F_{1}(W,Q^2)$ and $F_{2}(W,Q^2)$ inclusive structure functions are converted into the transverse $\sigma_{T}(W,Q^2)$ and the longitudinal $\sigma_{L}(W,Q^2)$ total cross section for the virtual-photon-proton interaction by using well-known relations between these quantities \cite{Ac16,Lin18,osip}:  
\begin{equation}\label{F1}
F_1= M_p\frac{K}{4\pi^2\alpha}\sigma_{T}(W,Q^2) \,,
\end{equation}
\begin{equation}\label{F2}
F_2= \nu\frac{\sigma_{T}(W,Q^2)+\sigma_{L}(W,Q^2)}{4\pi^2\alpha}\frac{(2\nu M_p-Q^2)Q^2}{2M_p(Q^2+\nu^2)} \,,
\end{equation}
where $M_p$ and $\nu$ stand for the proton mass and the energy transferred by the electron in the lab. frame (Lorentz invariant in electron scattering processes), respectively; $\alpha=\frac{1}{137}$, $K = \frac{2\nu M_p - Q^2}{2M_p}$.
The unpolarized inclusive virtual-photon-proton cross sections $\sigma(W,Q^2)$ are computed as:
\begin{equation}\label{sigma}
\sigma(W,Q^2)=\sigma_{T}(W,Q^2)+\varepsilon\sigma_{L}(W,Q^2) \,,
\end{equation}
\begin{equation}\label{eps}
\varepsilon=\Bigl(1+2\frac{\nu^2+Q^2}{Q^2}tg^2\frac{\theta_e}{2}\Bigl)^{-1},
\end{equation}
where $\theta_{e}$ is the electron scattering angle in the lab. frame
 
Inclusive electron scattering cross-sections are computed from the virtual-photon-proton  cross section:

\begin{equation}
    \frac{\mathrm{d}^2\sigma_{ep\to X}}{\mathrm{d}W \mathrm{d}Q^2}=\Gamma_{v} \sigma_{incl},
\end{equation}
where $\Gamma_{v}$ is virtual photon flux defined in the paper \cite{osip}.

\section{Results and discussion}
The inclusive electron scattering observables, the inclusive electron scattering and the total virtual photon-proton cross sections, the $F_{1}$ and $F_{2}$ 
structure functions can be evaluated in real time in the web-page \cite{Chesn19}. The computations can be done for the incoming electron beam energies, at the kinematics grid and for the particular observables requested by user. The evaluated observables can be produced both in the numerical format and as the graphical representations.

\begin{figure}
    \centering
    \includegraphics[width=6 cm]{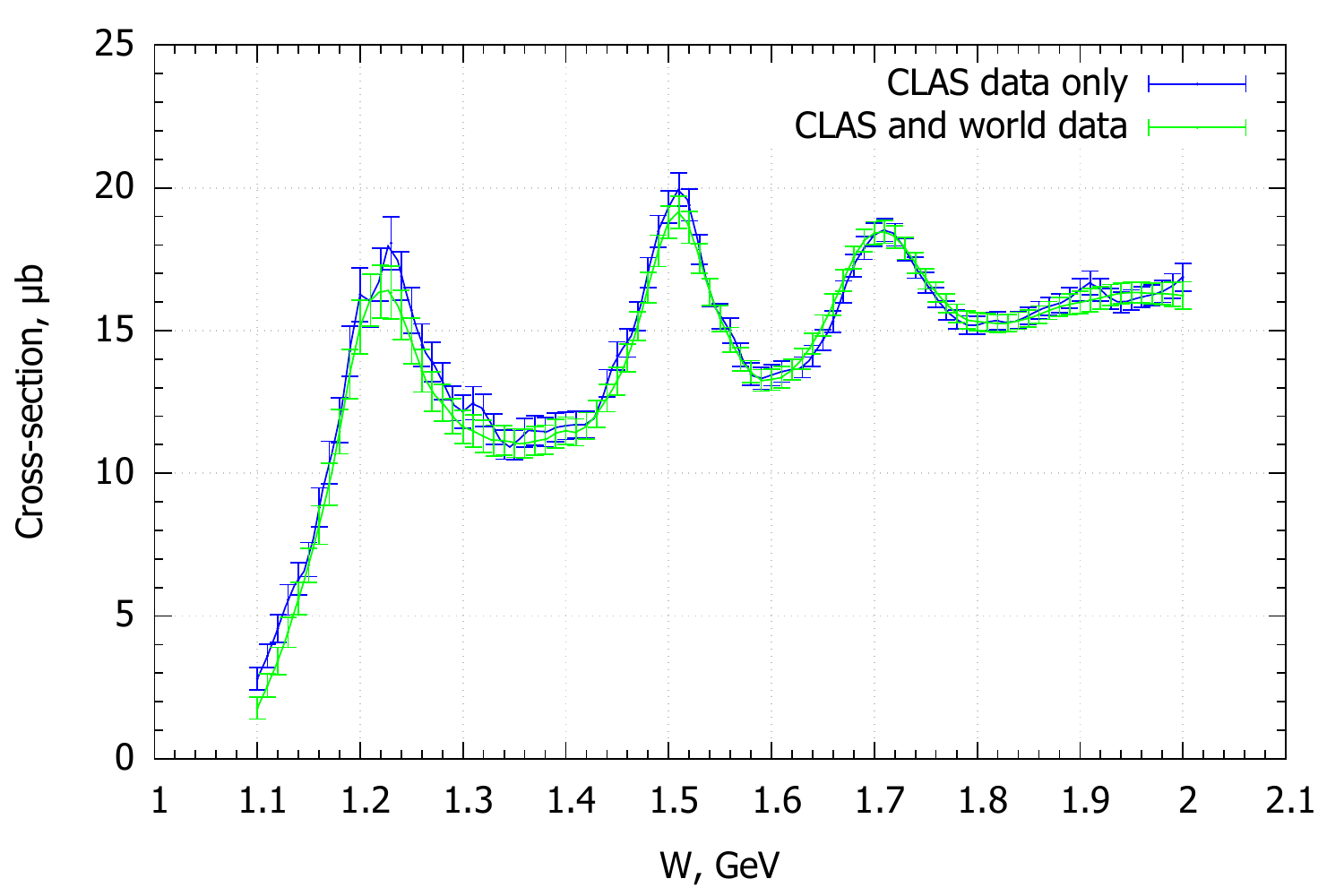}
    \includegraphics[width=6 cm]{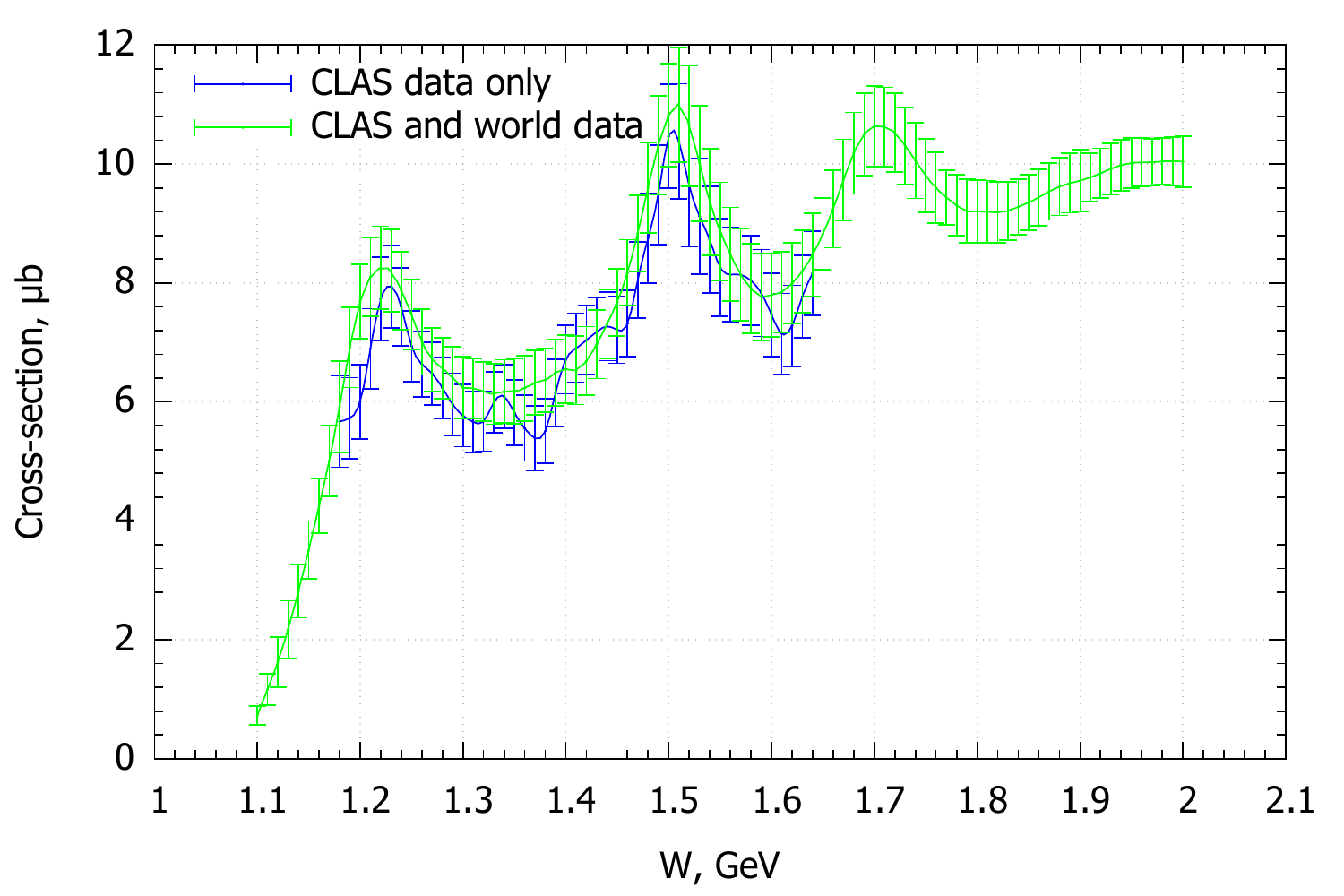}
    \caption{Comparison between total virtual-photon proton cross sections at Q$^2$=3.0 GeV$^2$ (left) and Q$^2$=4.0 GeV$^2$ (right) for electron beam energy $E_{e}=10.6$ GeV evaluated from the two sets of the experimental data: a) the CLAS results only \cite{osip} (blue) and b) CLAS and world data from the \cite{bost} parameterization (green). }
    \label{fig2}
\end{figure}

In Fig.~\ref{fig2} we compare unpolarized total virtual-photo-proton cross sections evaluated by using the two sets of the experimental data: a) the CLAS results only and b) the combination of the CLAS results and the world data. The comparison in Fig~\ref{fig2} are the representative examples. The results on total virtual-photon-proton cross sections evaluated from the CLAS data only and from the CLAS and world data parameterization \cite{bost} are consistent within the kinematics area covered by our approach 1.07 GeV $<$ W $<$ 4.0 GeV and 0.5 GeV$^2$ $<$ $Q^2$ $<$ 7.0 GeV$^2$ 

This consistency supports credible evaluation of the inclusive electron scattering observables within the framework of the presented in the proceeding procedure.

\begin{figure}
\centering
    \includegraphics[width=6 cm]{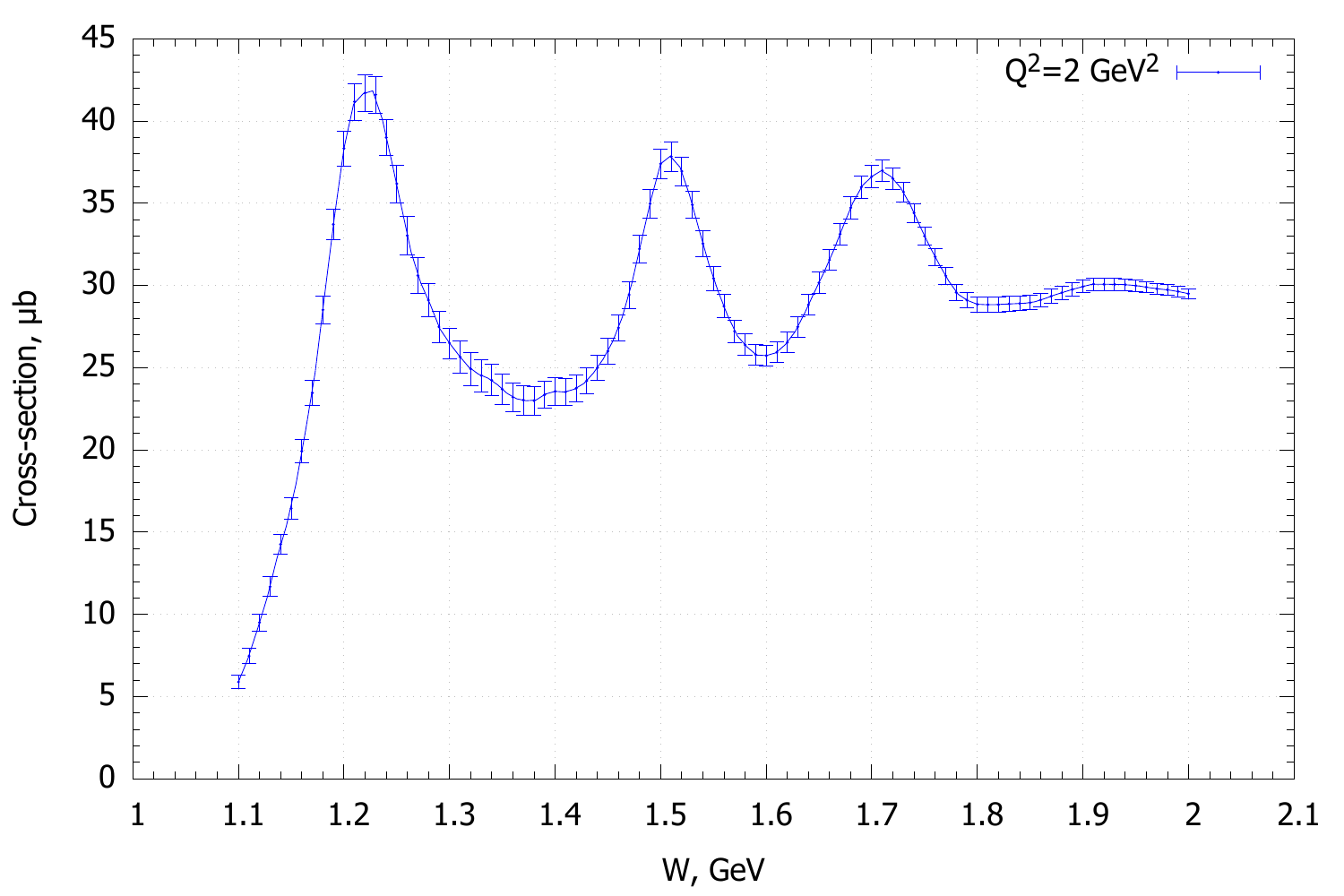}
    \includegraphics[width=6 cm]{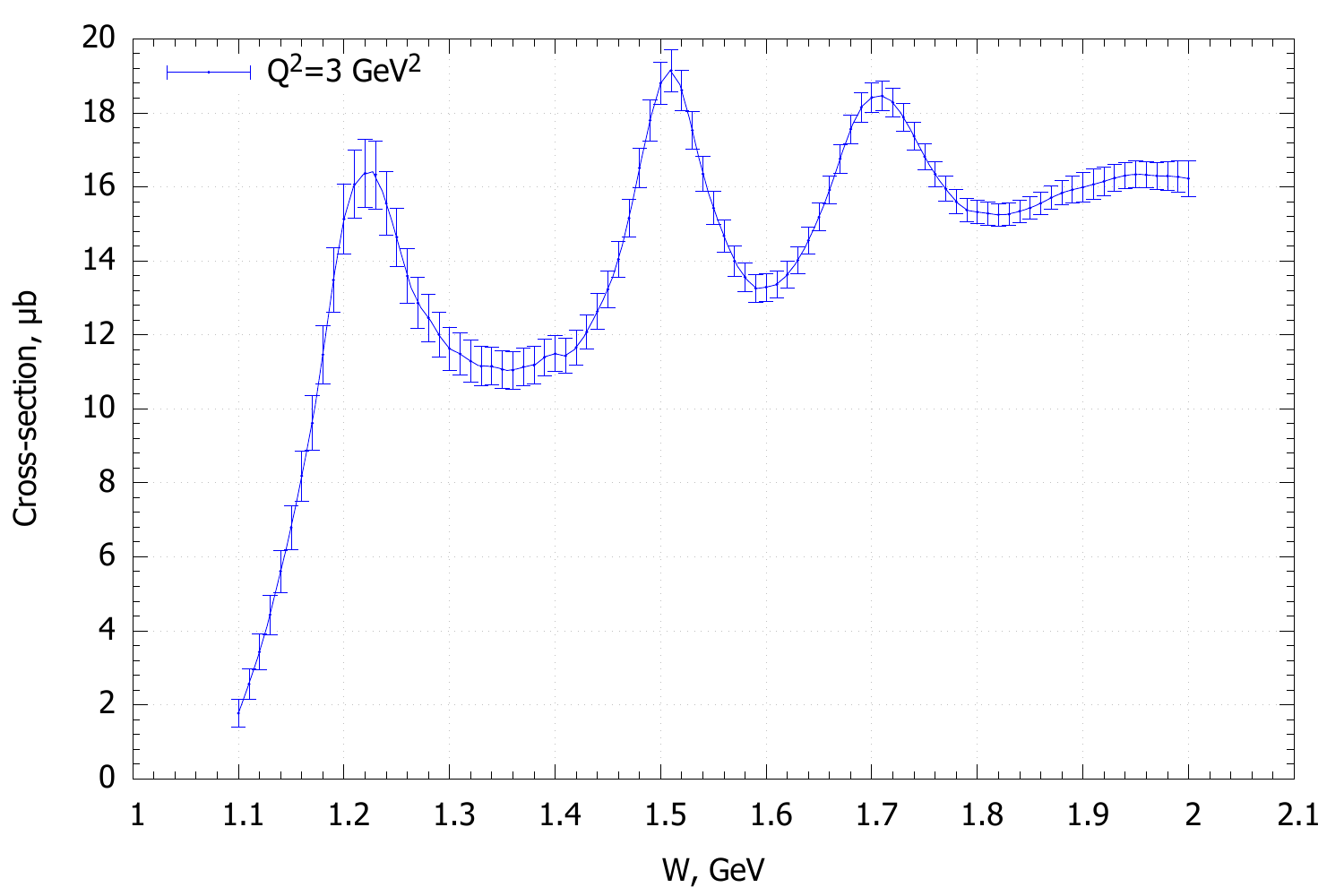}
    \includegraphics[width=6 cm]{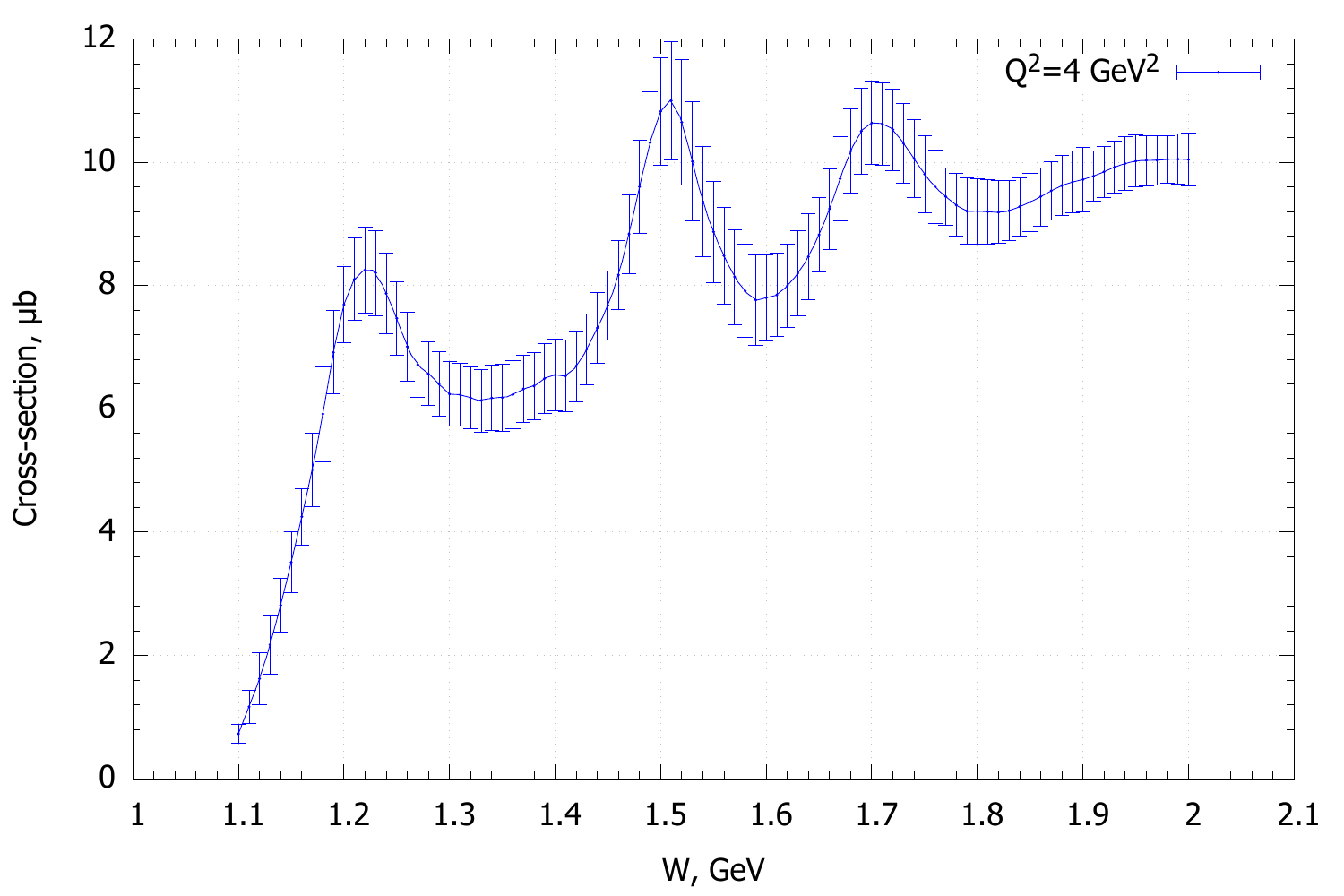}
    \includegraphics[width=6 cm]{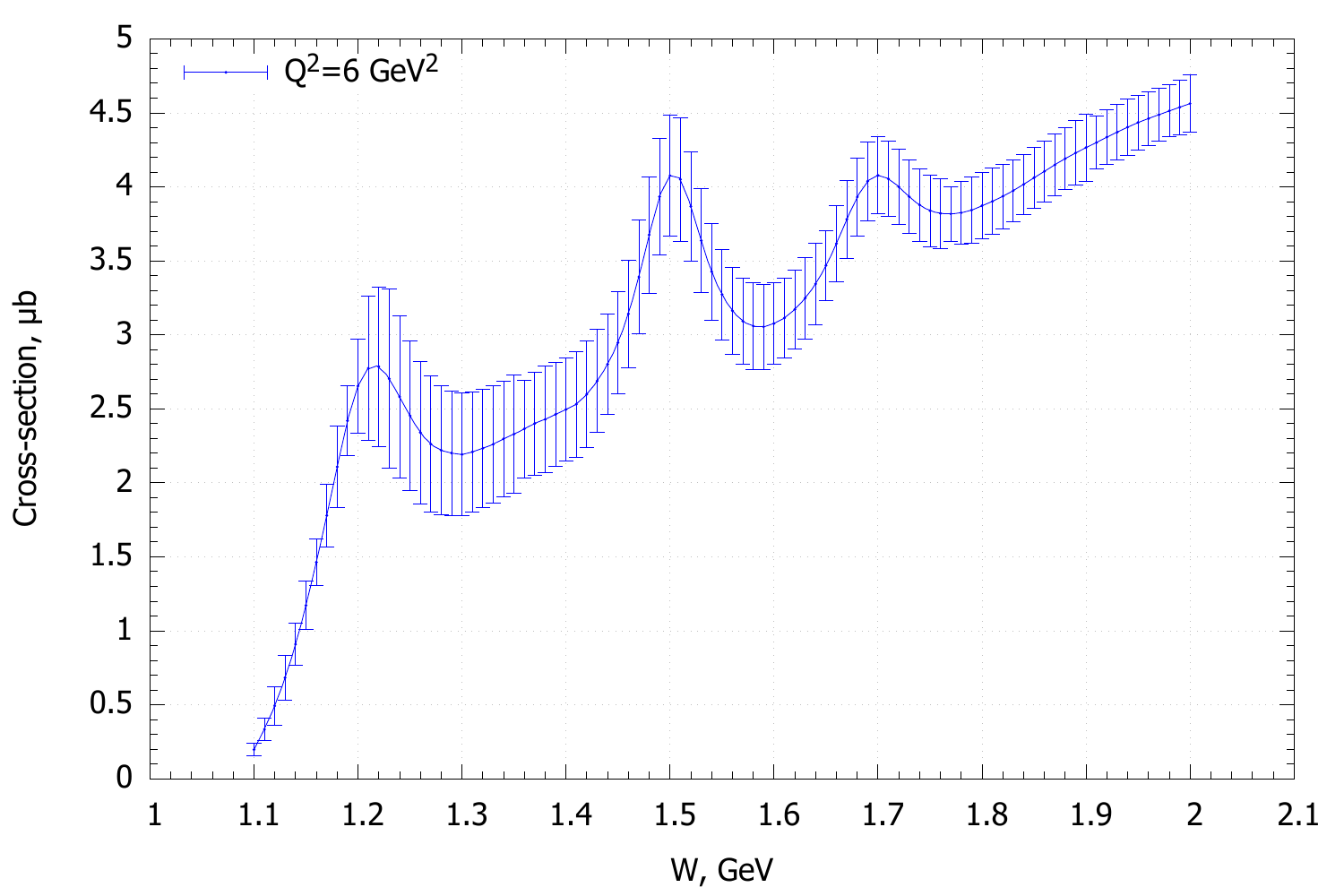}
    \caption{Evolution of the total virtual photon-proton cross sections in the resonance region with photon virtuality Q$^2$ for the incoming electron beam energy $E_{e}=10.6$ GeV}
    \label{fig3}
\end{figure}

The evolution of the total virtual-photon-proton cross section with the photon virtuality Q$^2$ estimated from the experimental data is shown in Fig~\ref{fig3}. 

The peaks from the first, second and third resonance regions are clearly seen demonstrating promising opportunity to extract $\gamma_{v}pN^*$-electrocouplings from exclusive meson electroproduction data with CLAS and the future data with the CLAS12 detector in the range of photon virtualities $Q^2$ $<$ 7.0 GeV$^2$.
Three resonant peaks demonstrate very different evolution with Q$^2$. The first peak consisted of the $\Delta(1232)3/2^+$ and the tail of $N(1440)1/2^+$ resonance contributions decreases with Q$^2$ most rapidly in comparison with two others peaks. Instead, the second resonance peak, composed by the contributions from $N(1440)1/2^+$, $N(1520)3/2^-$, and $N(1535)1/2^-$ resonances, demonstrates the most slow evolution with Q$^2$. Rate of the fall-off with Q$^2$ for the third resonance peak composed by several excited nucleon states with masses around 1.7 GeV is intermediate. This observation suggests distinctively different structural features for different excited nucleon states. Studies of $\gamma_{v}pN^*$-electrocouplings of all prominent resonances are needed in order to access many facets of strong QCD dynamics in the generation of full spectrum of the excited nucleon states.



\begin{figure}
    \centering
    \includegraphics[width=11.5 cm]{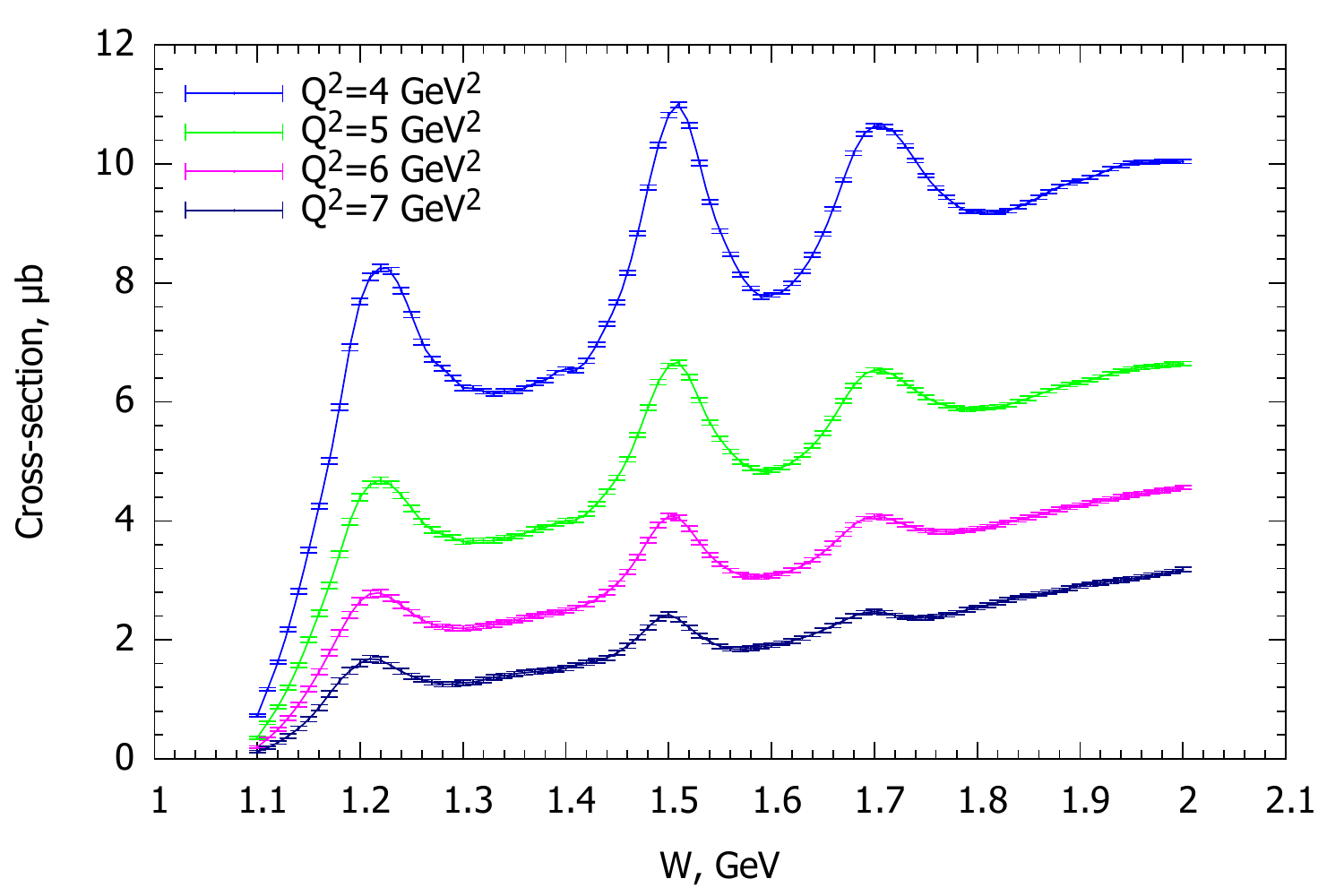}
    \caption{Projected CLAS12 results on the total virtual photon-proton cross sections at W $<$ 2.0 GeV and for the initial electron beam energy $E_{e}=10.6$ GeV. The error bars represent the statistical uncertainties evaluated for the integrated luminosity of 12.8*10$^{10}$ mcbn$^{-1}$ collected in the Spring 2018 run with the CLAS12.}
    \label{fig4}
\end{figure}

The CLAS12 is the only facility in the world capable to obtain the results on $\gamma_{v}pN^*$-resonance electrocouplings at the highest photon virtualities ever achieved in exclusive reactions of 5.0 GeV$^2$ $<$ Q$^2$ $<$ 12.0 GeV$^2$. The first results foreseen in these efforts will be the measurements of the inclusive structure function in the resonance region at high photon virtualities of Q$^2$ $>$ 5.0 GeV$^2$. We made prediction for the total virtual-photon-proton cross sections measured with the CLAS12 in the range of photon virtualities 4.0 GeV$^2$ $<$ Q$^2$ $<$ 7.0 GeV$^2$. The predicted cross section and their expected statistical uncertainties are shown in Fig~\ref{fig4}. The statistical uncertainties are evaluated for the integrated luminosity $12.8*10^{10}$ mcbn$^{-1}$ collected in the Spring 2018 run with the CLAS12 and for the bin sizes over W and Q$^2$ $\Delta W$=0.01 GeV and $\Delta Q^2$=0.1 GeV$^2$. 

The measurements with the CLAS12 detector will provide the results on inclusive electron scattering observables of highest statistical accuracy and within smallest W vs Q$^2$ bins ever achieved in the resonance region. The expected statistical uncertainties should be in the range from 0.2 \% to 2.0 \%. These experimental data will provide the information on the resonance contribution at high Q$^2$ $>$ 5.0 GeV$^2$ for the first time.

\section{Summary}

The method for evaluation of the inclusive electron scattering observables from the experimental data is developed allowing us to estimate the inclusive electron scattering, total virtual photon-proton cross sections and the $F_{1}$ and $F_{2}$ structure functions in the kinematics area of 1.07 GeV $<$ W $<$ 4.0 GeV and 0.5 GeV$^2$ $<$ Q$^2$ $<$ 7.0 GeV$^2$. The computations of these observables can be done in the real time at the kinematical grid defined by user in the interactive web page \cite{Chesn19}. The developed tool is of interest for analyses of the data from CLAS and the future data from the CLAS12. Estimated values of inclusive electron scattering observables allow us to check normalization of all semi-inclusive and exclusive processes under studies with the CLAS/CLAS12 and validate the credible evaluation of the electron detection efficiency used in the analyses of the experimental data. Projected inclusive electron scattering observables at Q$^2$ $>$ 5.0 GeV$^2$ are useful in the planning of the future experiments with the CLAS12 on exploration of hadron structure in the experiments with the electromagnetic probes.

\providecommand*{\BibDash}{}
\bibliographystyle{pepannt}
\bibliography{EMIN_biblio}

\end{document}